# Non-linear vector ANN predictor for Earth rotation parameters forecast

D.Milkov[1], L.Karimova[2], Z.Malkin[1]
[1]Central Astronomical Observatory at Pulkovo RAS
[2]Institute of Mathematics, Alma-Ata, Kazakhstan

e-mail: *d.a.milkov@gmail.com*

*Many approaches are developed for the forecasting of the Earth rotation parameters. In this work, we consider long-term vector prediction scheme realized on the artificial neural network. Learning set is formed on basis of the Taken' algorithm. Our approach allows us to obtain the vector of the parameter values and escape the exponential growth of the prediction errors. The versions of the prediction enhancement based on the using for nonlinear corrector are discussed.*

So far as Artificial Neuron Net (ANN) is representation "in-out" signal and entering vector is distributed in multidimensional space of attributes we realized the procedure of learning table building for which the main constructing concepts bases on the dynamic features. In this case the space of attributes transforms qualitative into phase space of dynamic system which, as we suppose, generates the concerned time series[1-5]. The ANN realizes (or very close approximates) representation in the space. Correspondingly it is structural changed the representation of probability measure with a glance dynamic system behavior[6]. And the learning vector becomes the points of phase space which belongs the phase system trajectory.

The suggest procedure of table building for learning of ANN bases on famous Takens theorem. Considering time series as the foot of phase trajectory of dynamic system in unrestricted direction on real axis this theorem claims the time series representation in the space of required dimension exists. And this representation is typically embedding. In addition, obtaining geometrical image is the topological copy of original attractor of the dynamic system.

The simplified version of Takens theorem and algorithm of the time series reconstruction dynamics may be represent the following. Suppose the k-dimensional state vector $\mathbf{x_t}$ evolves under unknown but continuous determine dynamics such as the diffeomorphism group in the phase space $R^k$. Let the system attractor is compact d-dimensional manifold. Suppose the one-dimensional observable $y_t : R^k \to R$ is smooth function which related to all components of this vector[6,7].

In every moment of the time we have the current value $y_t$ and the preceding time moment observations which is multiple certain time lag $\tau$: $y_{t-\tau}$, $y_{t-2\tau}$, $y_{t-3\tau} \cdots$ The number of usable past counts are constantly and equally to the dimension of embedding. Therefore, using k time lags we obtain the k-vector. The sequence of such vectors which is generated by shifts of k-blocks in the line of time series forms the model trajectory. We hope the movement in this space will be more predictable and probably in limit $k \to \infty$ becomes determine. In fact, the dynamics of such vectors becomes determine at finite *k*. apparently the such trajectory will be converge on certain compact i.e. model in own space.

It is claimed that obtaining the required construction by embedding we may numerically investigate many features of the dynamics. For example, the attractor dimension, Lyapunov exponents, displaying the entry condition sensitivity degree, and multidimensional analog of autocorrelation function named the autocorrelation dimension.

So, this method of the building of learning set was developed for prediction using ANN where the main destination is modeling the asymptotic behavior of phase trajectory and extrapolation in time, i.e. prediction service.

The building of learning set consists in several stages.

The first consists in estimation of the dynamic features of the considering time series. In the first stage we obtain the emending dimension and time lag[8,9].

In the second stage we construct the vectors using method which is described above. The obtaining vectors is filled the rows in learning table of ANN step-by-step in compliance with movement in the line of time series[8,9].

The application of this method leads to interesting effect. The ANN which is learned on such sample realized the vector scheme of prediction. Contrast to one-step scheme of prediction the error of prediction is not accumulated through iterative using prediction-following counts.

In this work we used the following scheme of ANN:

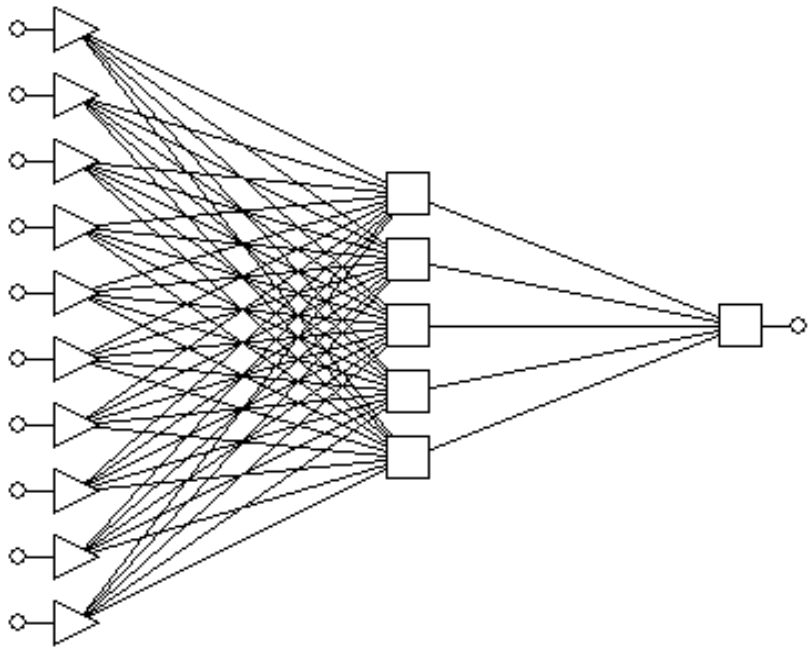

The size of embedding is $m=10$. For 10, 100, 200, 300 day forecast the time lag was correspondingly $\tau=10, \tau=100, \tau=200, \tau=300$.

It was done several forecast series. For every forecast (10, 100, 200, 300 days) ANN started $k=10$ times.

Then we moved back on the time line through 30 days. And so on in the past altogether was $n=11$.

The 100, 200, 300 days forecast was done for $X_p$, $Y_p$, UT1. The 10 day forecast was done only for $X_p$.

For example, in Figs. 1-39, the results of different day forecasts (modulus errors) from different points of the time are shown. One curve corresponds to one start of ANN.

Figures 40-42 present the RMS and MAE statistics:

$$RMS = \sqrt{\frac{\sum_{i=1}^{n}(forecast\_average_i - real\_value_i)^2}{n-1}}$$

$$MAE = \frac{\sum_{i=1}^{n}|forecast\_average_i - real\_value_i|}{n}$$

From our results we see that it is difficult to draw a final conclusion about quality of our forecast. It needs more detailed statistics on greater sample.

However, we can say that 200 day forecast results are very interesting between 100 and 200 days. Especially if we compare with [10].

**References**

1. Ott E., Sauer T., Yorke J.A. Coping with Chaos, J.Wiley & Sons, 1994
2. Chaos and Forecasting //Philosophical Transaction, v.348, N 1688, 1994, p.323-538
3. Abarbanel H.D.I., Carroll T.A., Pecora L.M., Sidorowich J.J., Tsimring L.S. Predicting physical variables in time-delay embedding. // Physical review E, V.49, N. 3, march 1994.
4. Данилкина Е.Б., Куандыков Е.Б., Каримова Л.М., Макаренко Н.Г. Методы топологического вложения в нейропрогнозе финансовых временных рядов// Нейроинформатика-2001, ч.2., Москва, 2001, с.13-27
5. Tribelsky M., Harada Y., Kuandykov Y., Makarenko N. Predictability of Market Prices// In Empirical Science of Financial Fluctuations: The Advent of Econo-physics, Springer-Verlag, Tokyo, 2002, p.241-249
6. Sauer T., Yorke J.A., Casdagli M. Embedology// J.Statist.Phys. v.65., N3/4, 1991, p.579-616
7. Rapp P.E., Schmah T.I., Mees A.I. Models of knowing and yhe investigation of dynamical systems. // Physica D 132 (1999) 133-149
8. Ott E., Sauer T., Yorke J.A. Coping with chaos: Analysis of chaotic data and the exploitation of chaotic systems // John Wiley and Sons. – 1994. – 432 p.
9. Parker, T. S., Chua, L. O. Practical Numerical Algorithms for Chaotic Systems. Springer, 1989. – 348 p.
10. http://www.cbk.waw.pl/EOP_PCC/

**$X_p$, forecast 10, errors, arcsec**

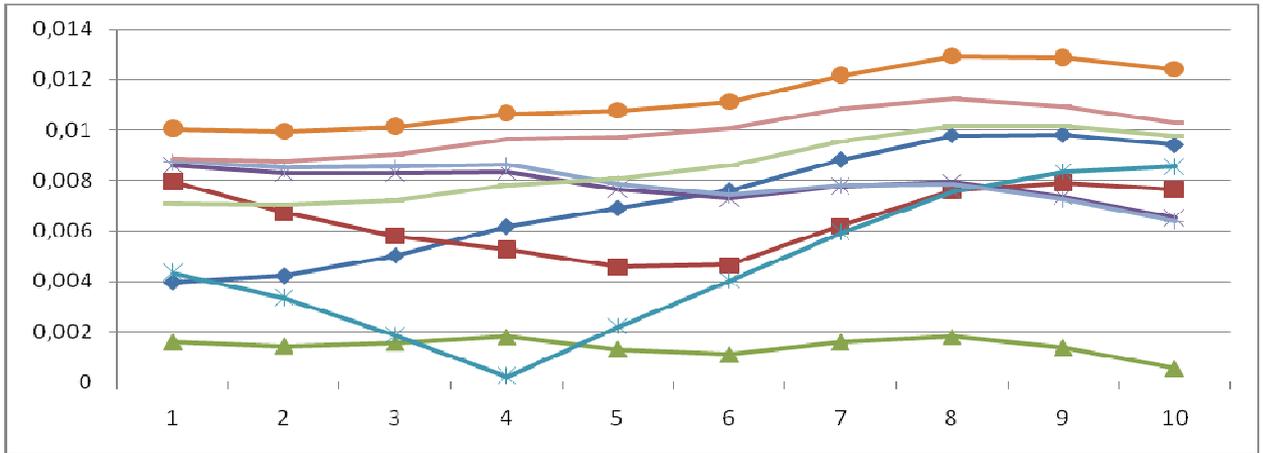
Figure 1 ($T_0$=20070902)

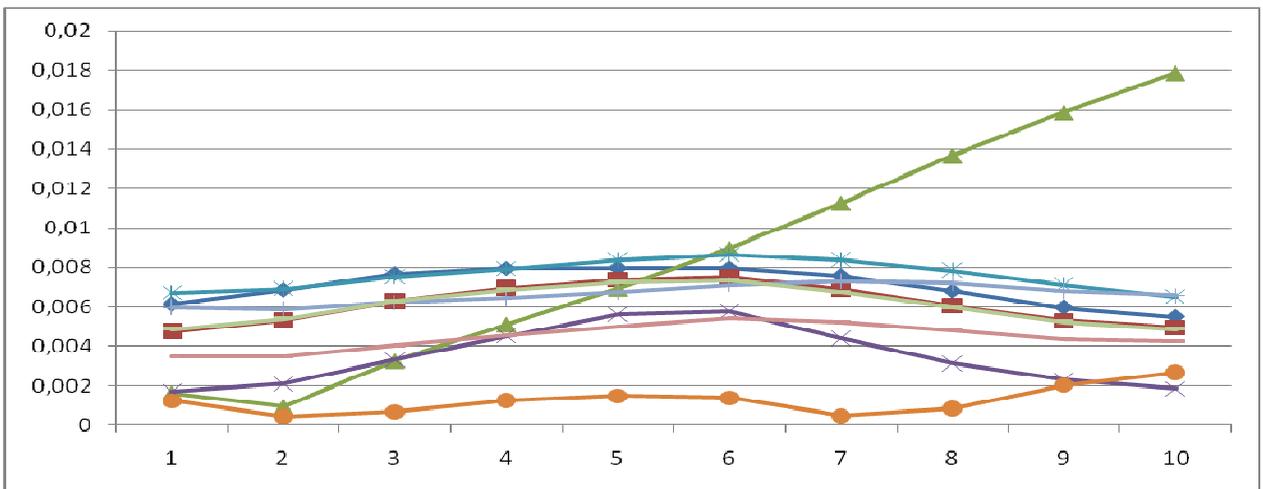
Figure 2 ($T_0$=20070804)

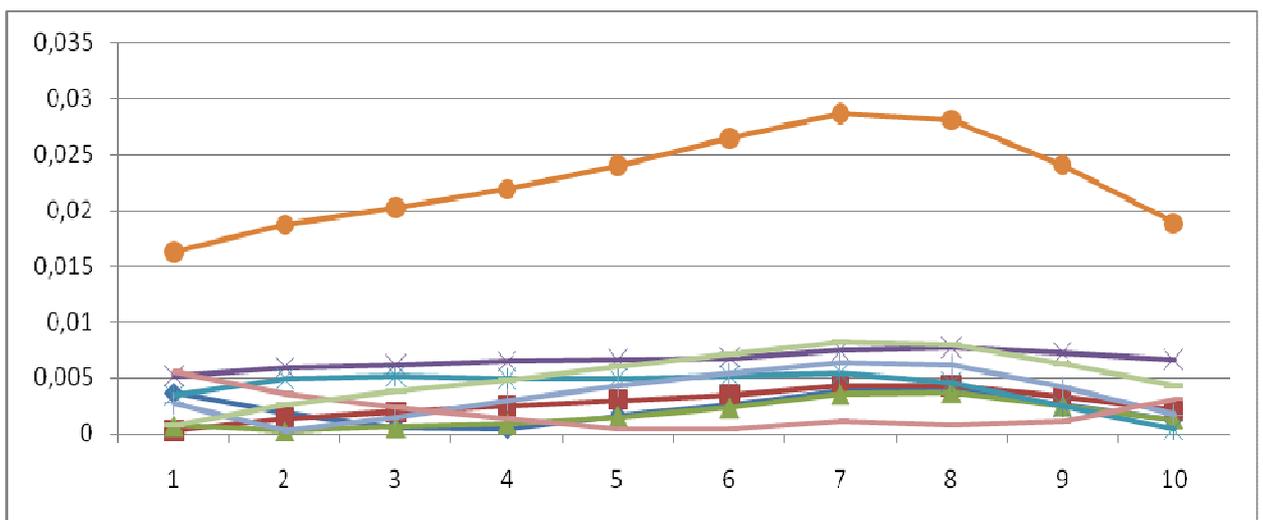
Figure 3 ($T_0$=20061107)

**$X_p$, forecast 100, errors, arcsec**

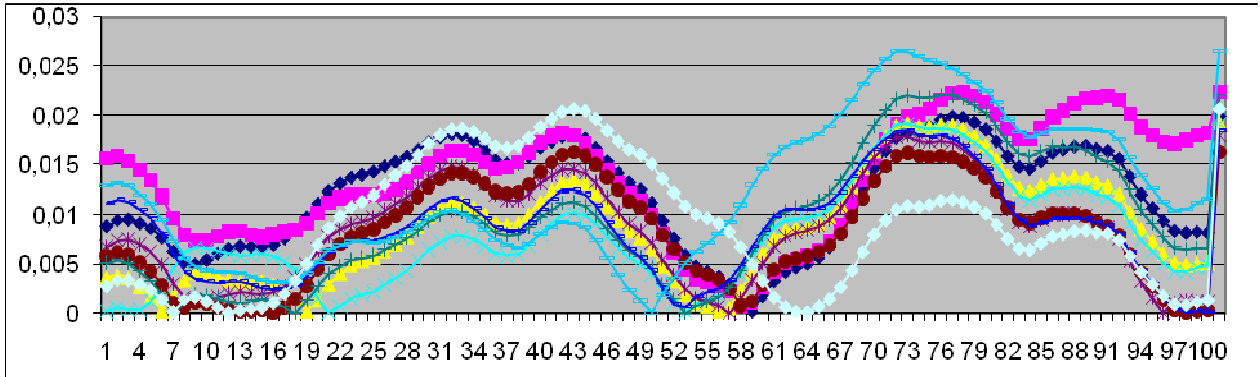

Figure 4 ($T_0$=20070902)

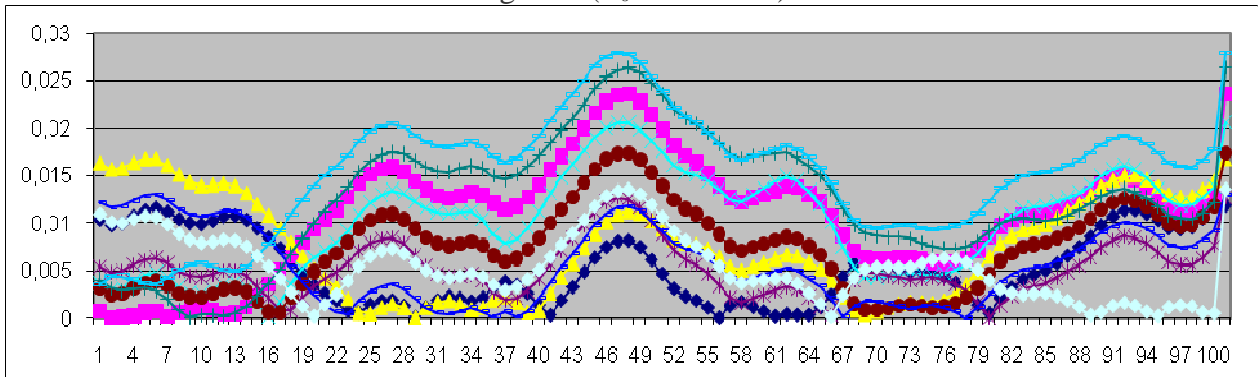

Figure 5 ($T_0$=20070804)

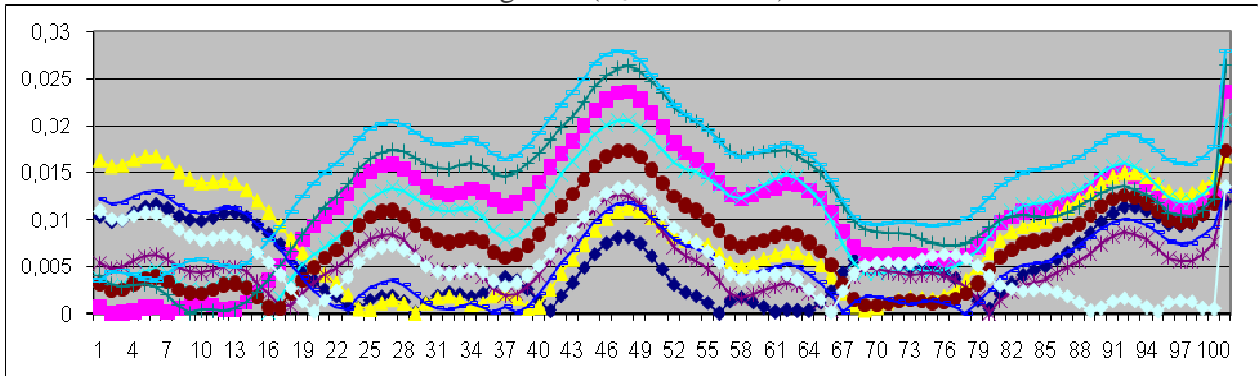

Figure 6 ($T_0$=20070406)

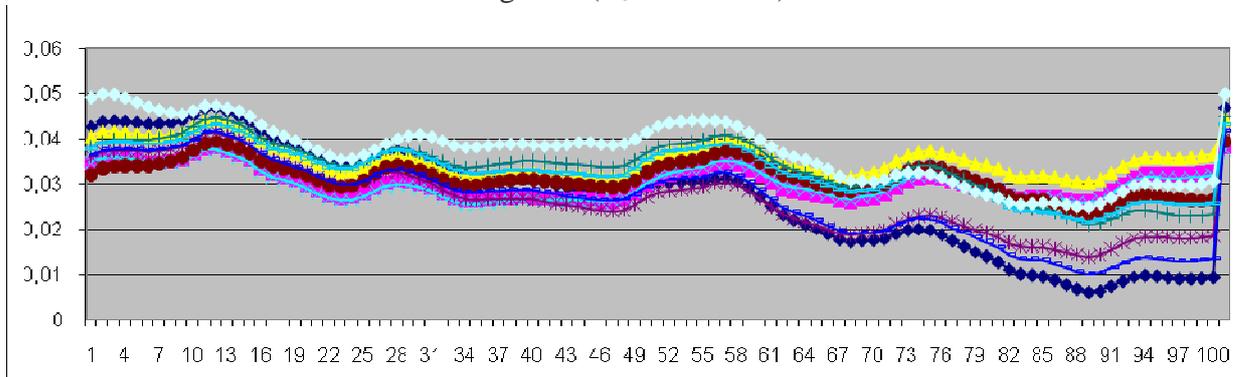

Figure 7 ($T_0$=20061107)

**X$_p$, forecast 200, errors, arcsec**

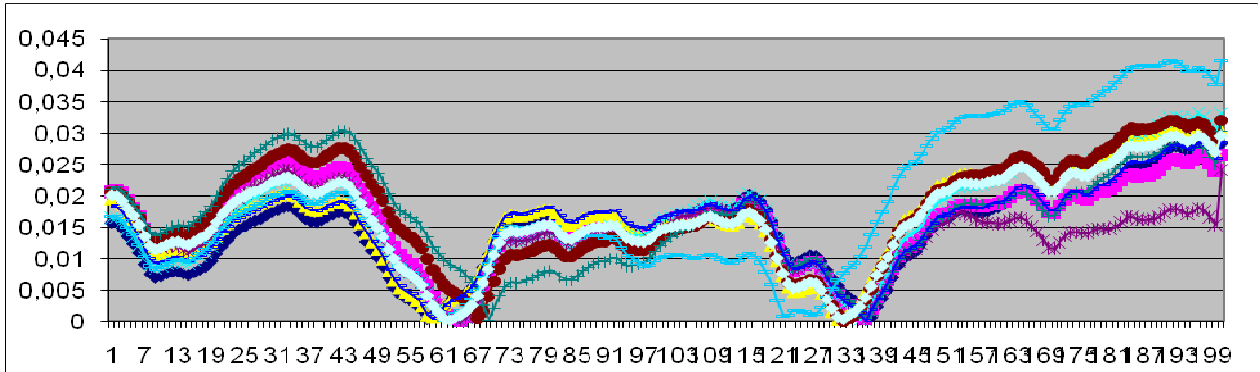

Figure 8 (T$_0$=20070902)

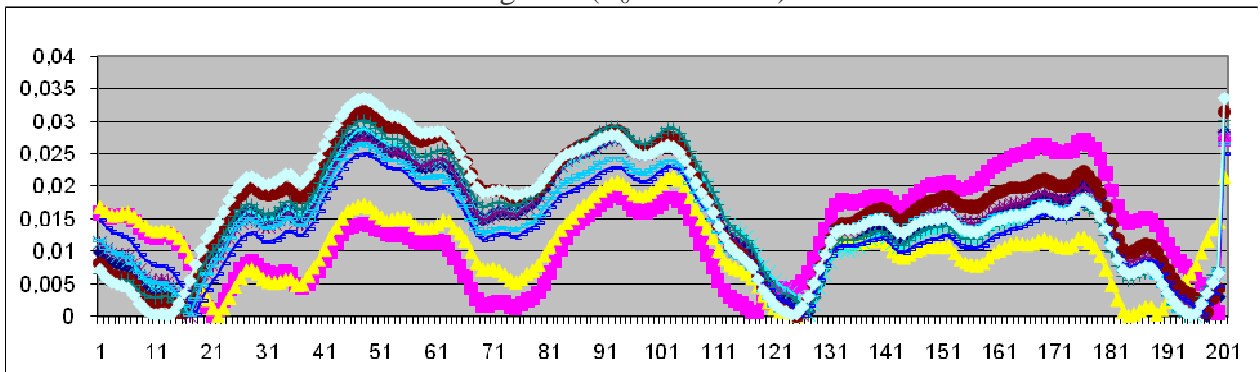

Figure 9 (T$_0$=20070804)

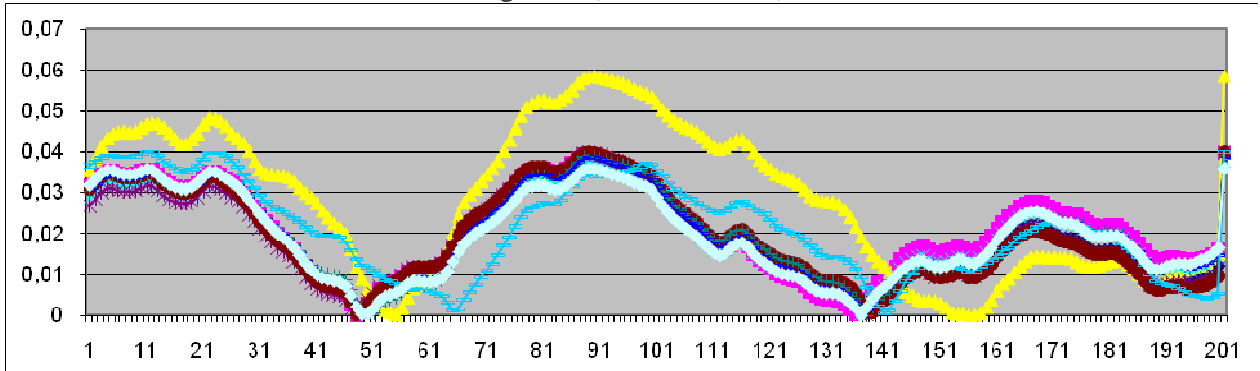

Figure 10 (T$_0$=20070406)

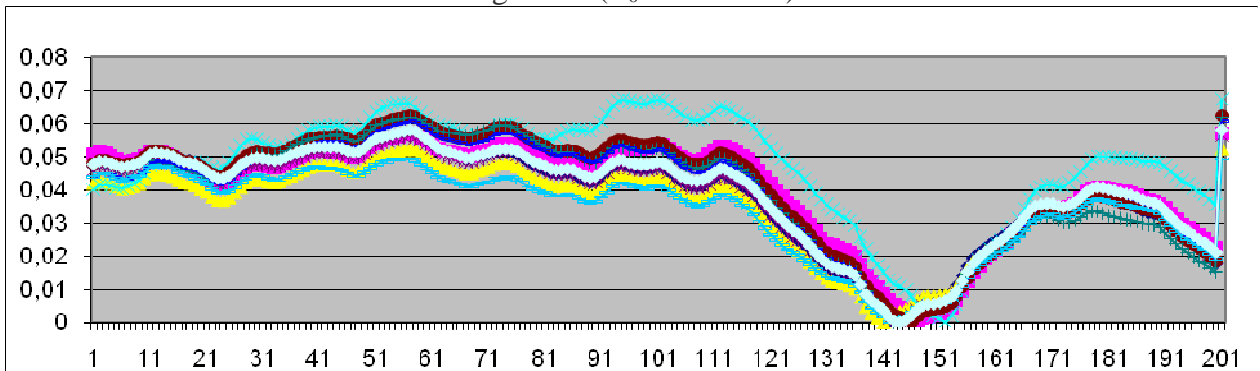

Figure 11 (T$_0$=20061107)

**X$_p$, forecast 300, errors, arcsec**

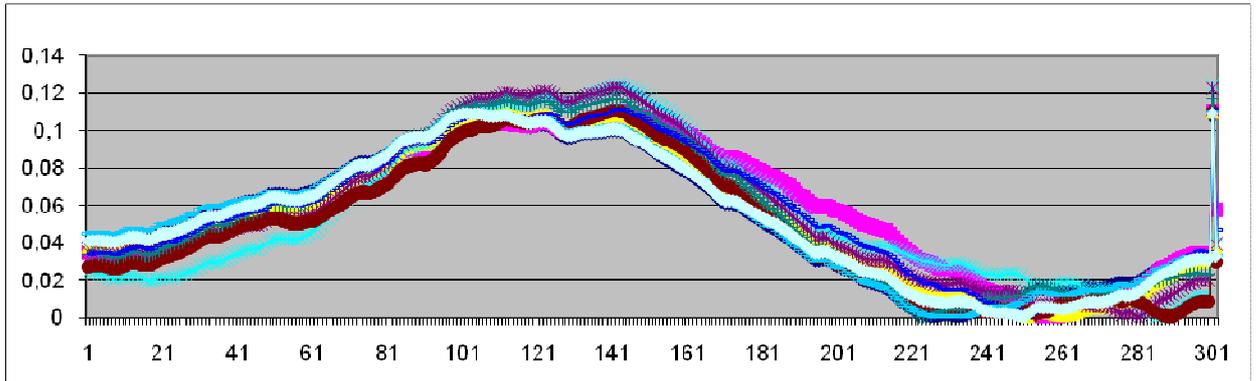

Figure 12 (T$_0$=20070902)

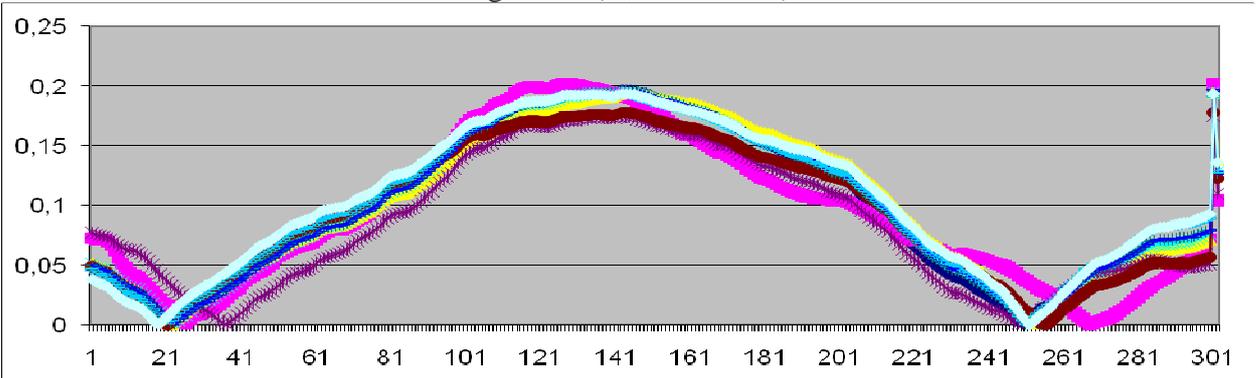

Figure 13 (T$_0$=20070804)

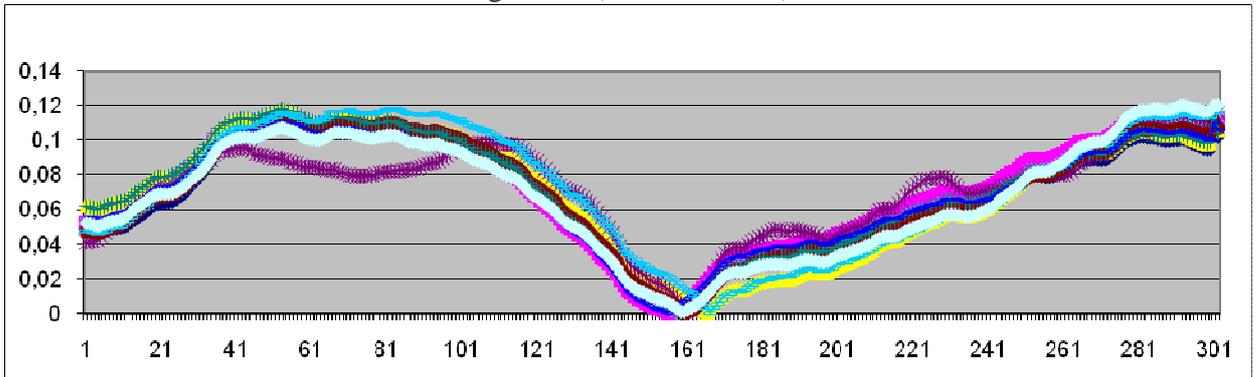

Figure 14 (T$_0$=20070406)

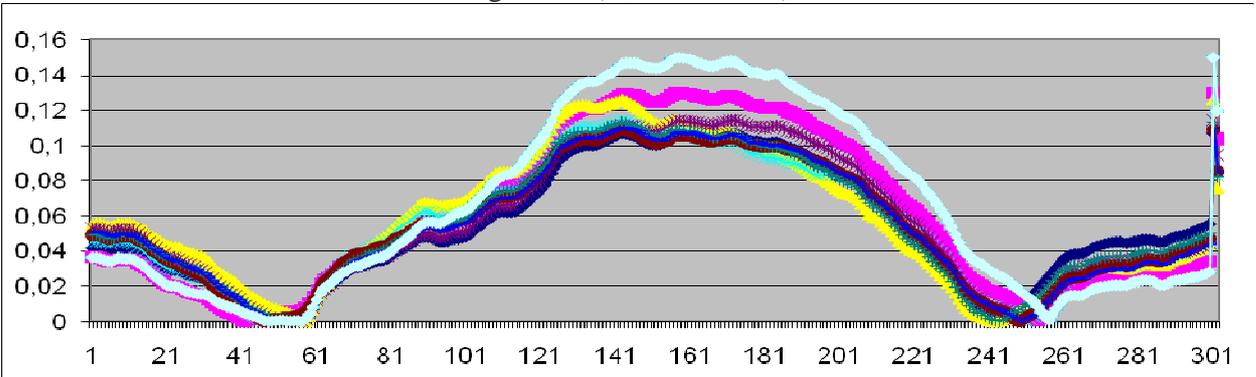

Figure 15 (T$_0$=20061107)

**Y$_p$, forecast 100, errors, arcsec**

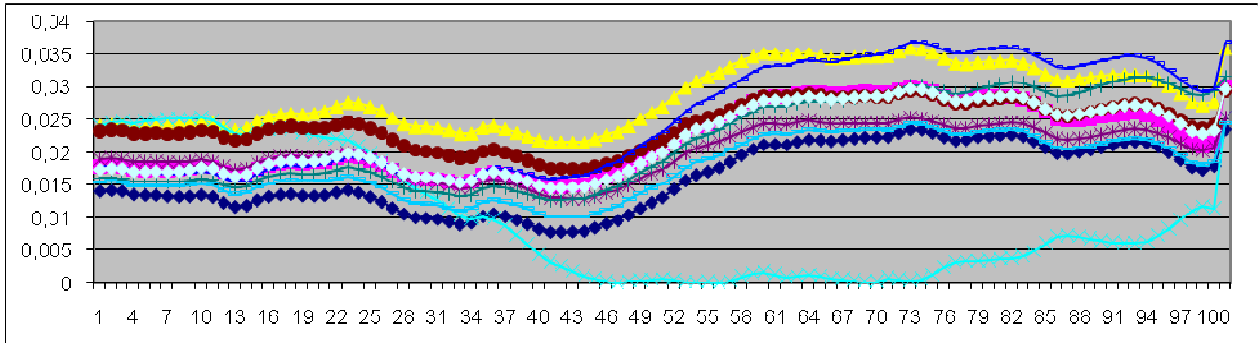

Figure 16 (T$_0$=20070902)

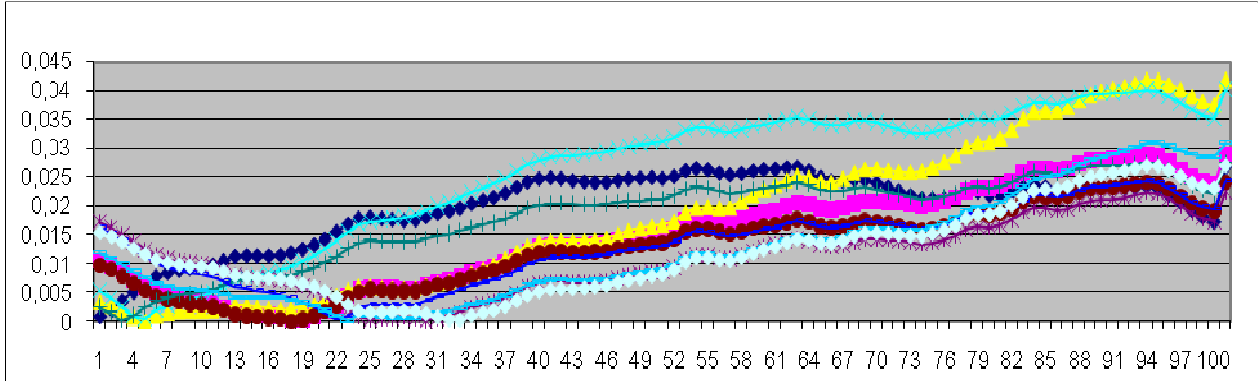

Figure 17 (T$_0$=20070804)

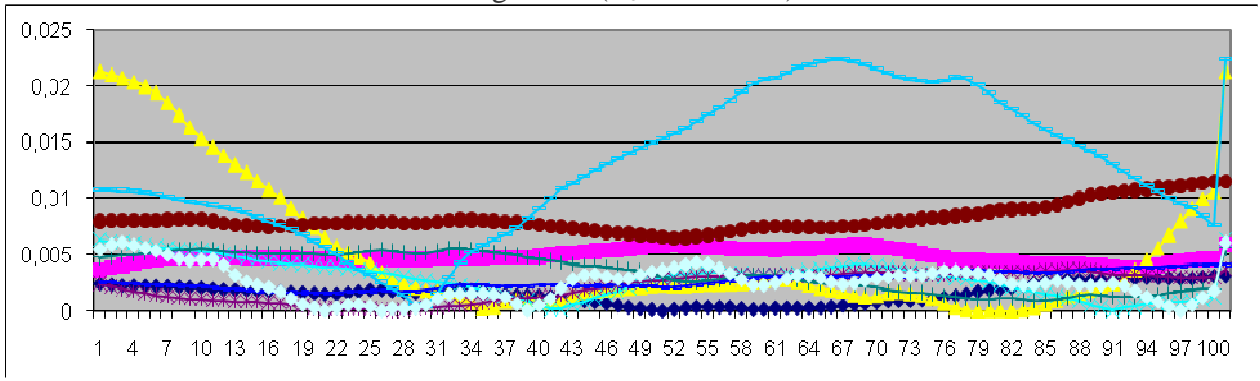

Figure 18 (T$_0$=20070406)

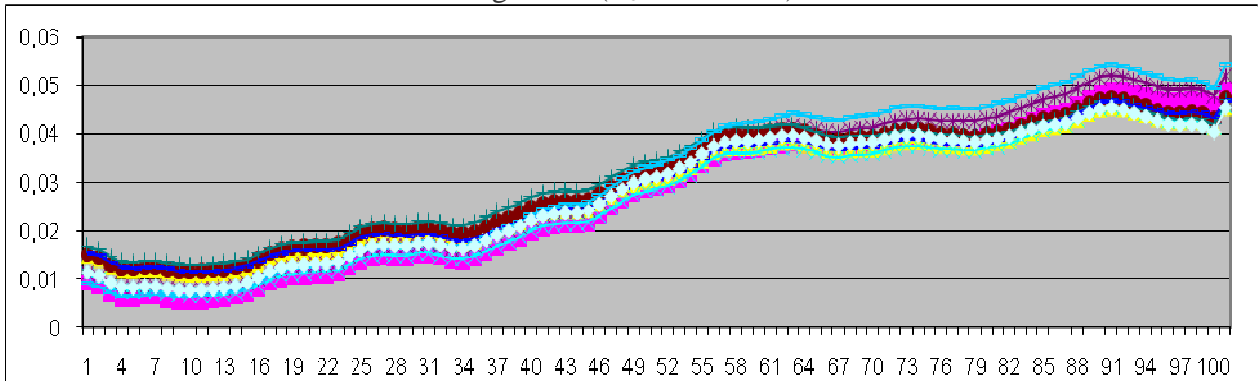

Figure 19 (T$_0$=20061107)

**$Y_p$, forecast 200, errors, arcsec**

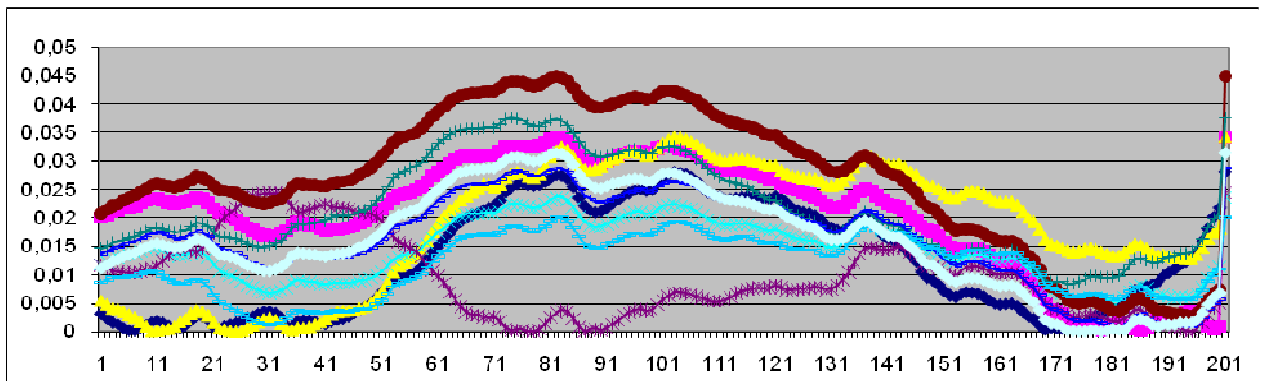

Figure 20 ($T_0$=20070902)

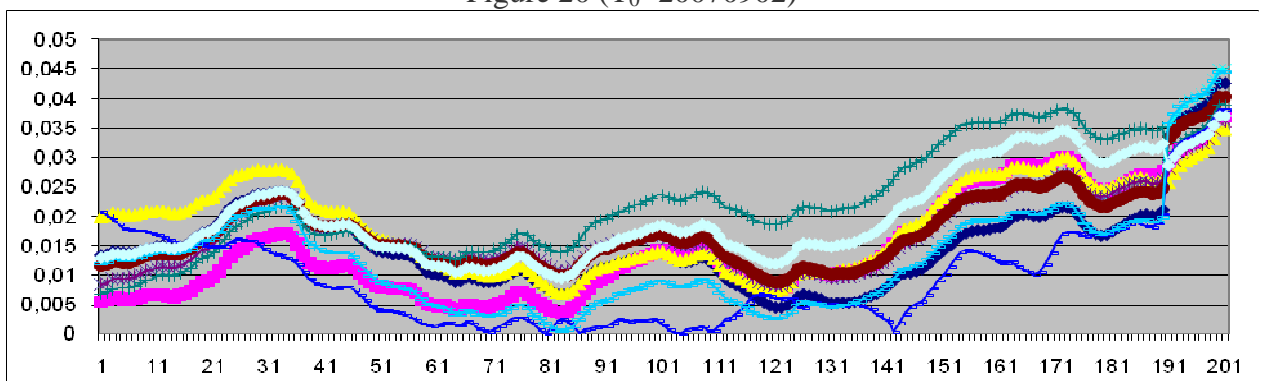

Figure 21 ($T_0$=20070804)

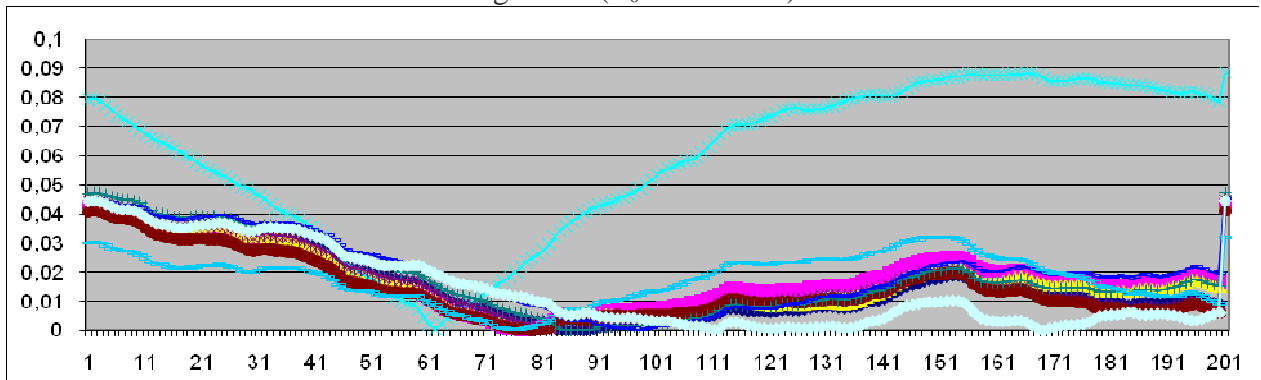

Figure 22 ($T_0$=20070406)

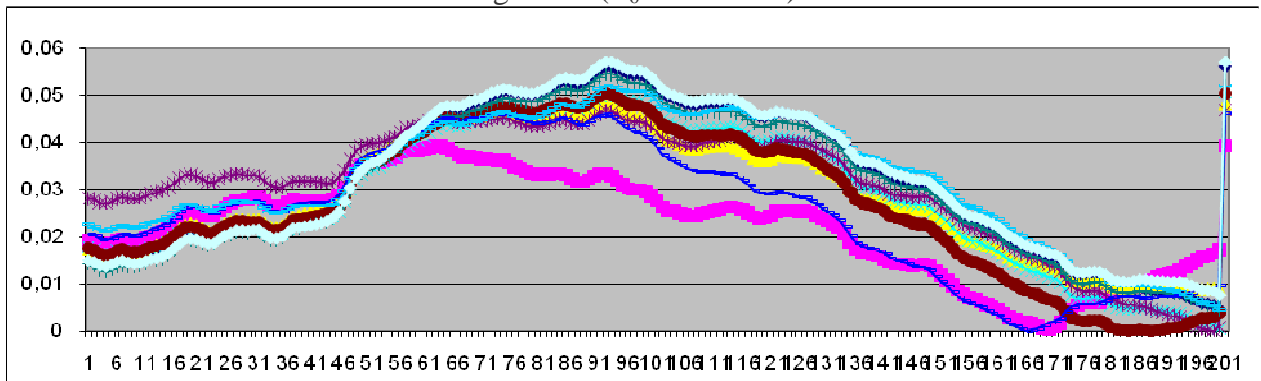

Figure 23 ($T_0$=20061107)

**$Y_p$, forecast 300, errors, arcsec**

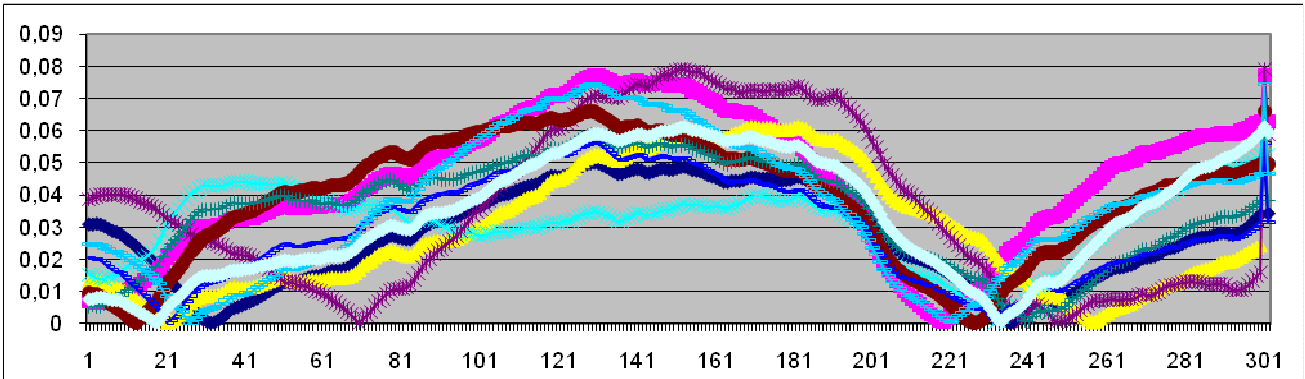

Figure 24 ($T_0$=20070902)

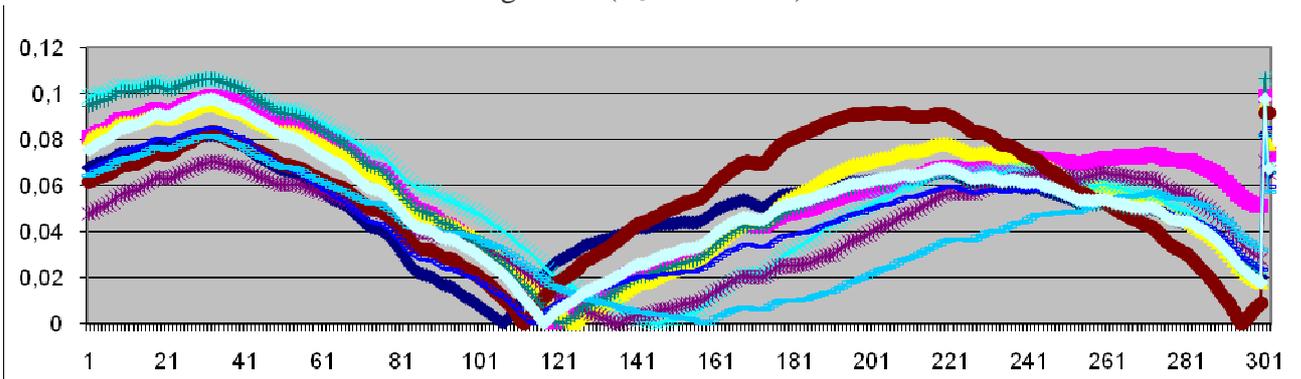

Figure 25 ($T_0$=20070804)

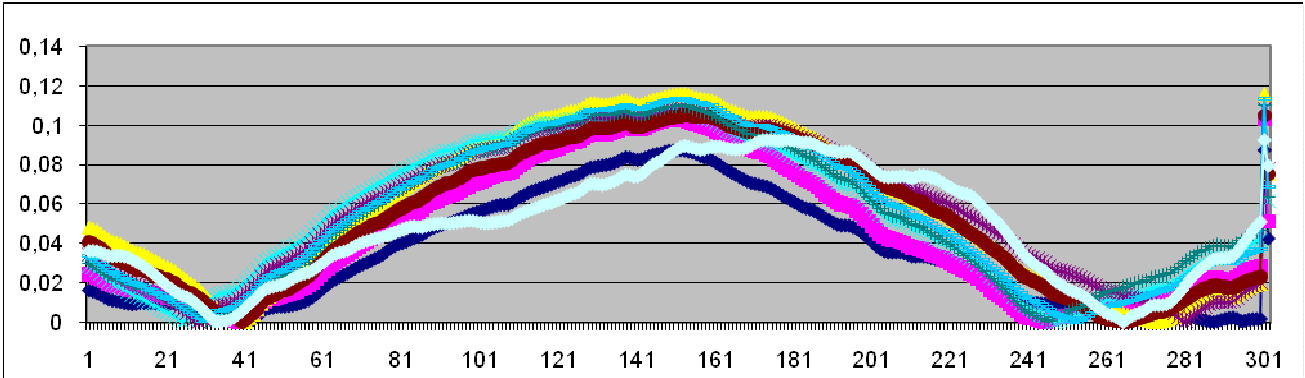

Figure 26 ($T_0$=20070406)

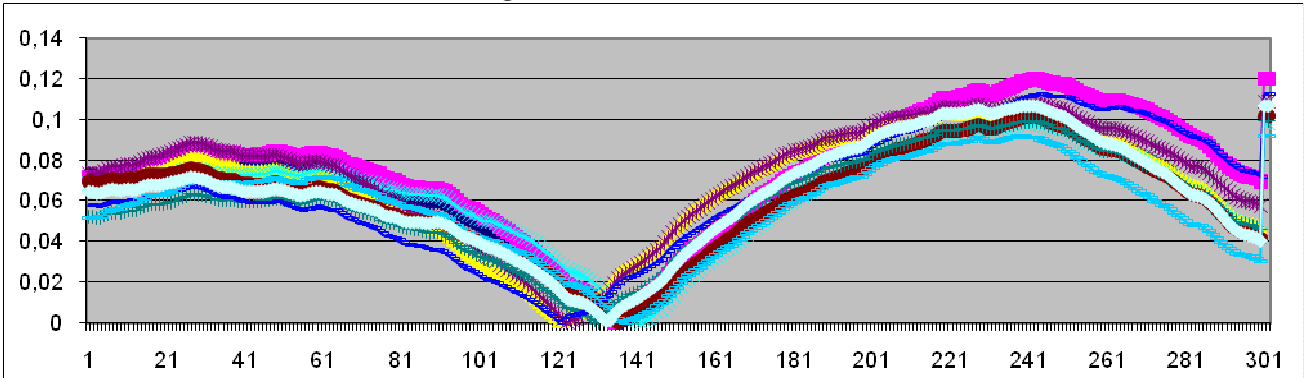

Figure 27 ($T_0$=20061107)

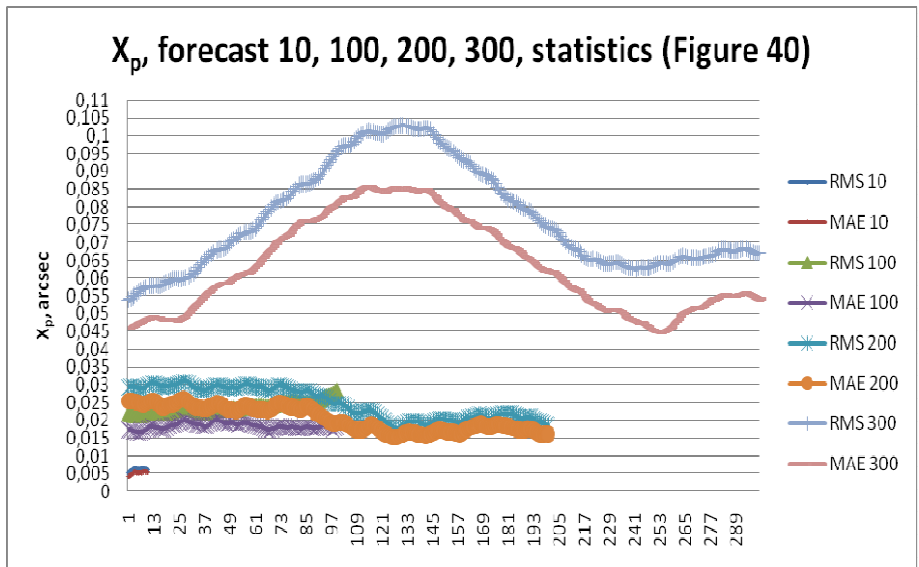

Figure 40. RMS and MAE for Xp forecast.

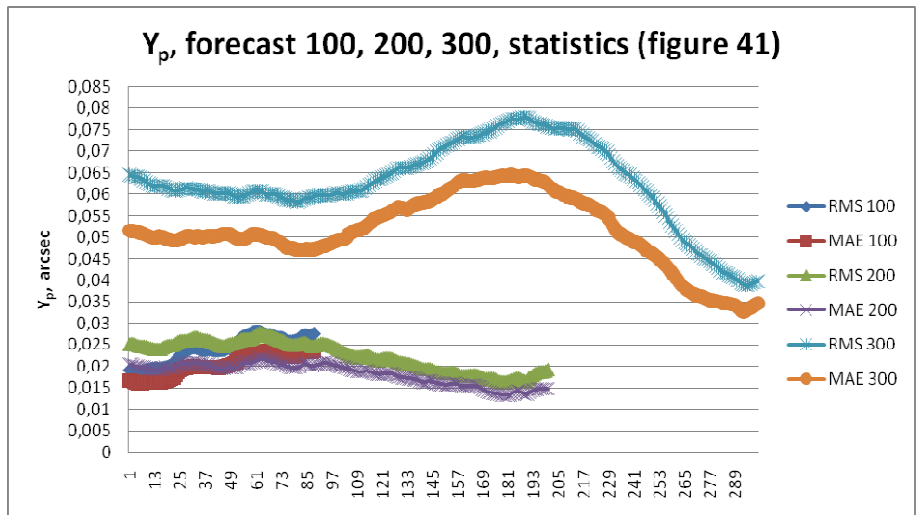

Figure 41. RMS and MAE for Yp forecast.

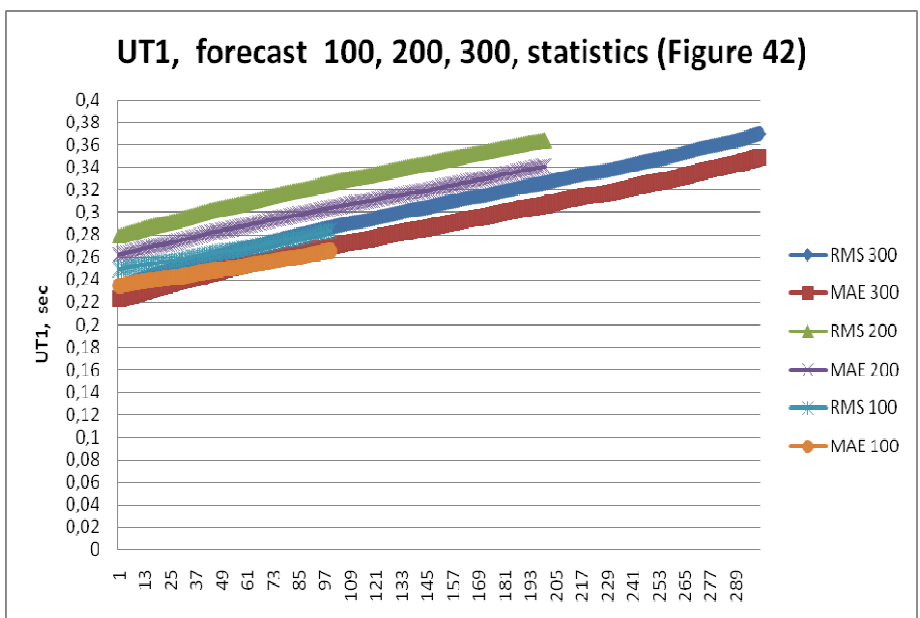

Figure 42. RMS and MAE for UT1 forecast